# A nonreciprocal optical resonator with broken time-invariance for arbitrarily high time-bandwidth performance

Ivan Cardea[1], Davide Grassani[1,2], Simon J. Fabbri[1], Jeremy Upham[3], Robert W. Boyd[3], Hatice Altug[4], Sebastian A. Schulz[5], Kosmas L. Tsakmakidis[6], Camille-Sophie Brès[1*].

[1]Ecole Polytechnique Fédérale de Lausanne (EPFL), Photonic Systems Laboratory (PHOSL), Lausanne CH-1015, Switzerland.

[2]Currently with Dipartimento di Fisica, Università degli studi di Pavia, via Bassi 6, 27100 Pavia, Italy.

[3]Department of Physics, University of Ottawa, Ottawa, ON, Canada.

[4]Ecole Polytechnique Fédérale de Lausanne (EPFL), Bionanophotonics System Laboratory (BIOS), Lausanne CH-1015, Switzerland.

[5]School of Physics and Astronomy, SUPA, University of St Andrews, St Andrews, KY169SS, UK.

[6]Solid State Physics section, Department of Physics, National and Kapodistrian University of Athens, Panepistimioupolis, GR - 157 84, Athens, Greece.

*Correspondence to: camille.bres@epfl.ch.

Most present-day resonant systems, throughout physics and engineering, are characterized by a strict time-reversal symmetry between the rates of energy coupled in and out of the system, which leads to a trade-off between how long a wave can be stored in the system and the system's bandwidth. Any attempt to reduce the losses of the resonant system, and hence store a (mechanical, acoustic, electronic, optical, atomic, or of any other nature) wave for more time, will inevitably also reduce the bandwidth of the system. Until recently, this time-bandwidth limit has been considered fundamental, arising from basic Fourier reciprocity. A recent theory suggested that it might in fact be overcome by breaking Lorentz reciprocity in the resonant system, reinvigorated a debate about whether (or not) this was indeed the case. Here, we report an experimental realization of a cavity where, inducing nonreciprocity by breaking the time invariance, we do overcome the 'fundamental' time-bandwidth limit of ordinary resonant systems by a factor of 30, in full agreement with accompanying numerical simulations. We show that, although in practice experimental constraints limit our scheme, the time bandwidth product can be arbitrarily large, simply dictated by the finesse of the cavity. Our experimental realization uses a simple macroscopic, time-variant, fiber-optic cavity, where we break Lorentz reciprocity by non-adiabatically opening the cavity, injecting a pulse of large bandwidth, and then closing the cavity, storing the pulse which can be released on-demand at a later time. Our results open the path for designing resonant systems, ubiquitous in physics and engineering, that can simultaneously be broadband (i.e., ultrafast) and possessing long storage times, thereby unleashing fundamentally new functionalities in wave physics and wave-matter interactions.

The time-bandwidth product (TBP) is a relational property characterizing all individual resonators, whether they are of mechanical, acoustic, electrical, atomic or optical nature. A general definition of the TBP should consider the product between the *acceptance* bandwidth ($\Delta\omega_{acc}$) of the system, which does not necessarily coincide with the measured cavity linewidth, as will be explained later, and its characteristic decay ('storage') time ($\tau_{out}$). The majority of present-day resonant systems, are *reciprocal* in nature and, consequently, time-reversal symmetric. Therefore, if wave energy may be coupled out of such systems, an exactly equal amount of energy can be coupled into them simply by 'reversing' time. In practice, in a reciprocal resonant system, $\Delta\omega_{acc}$ coincides with the cavity linewidth, and, therefore, its TBP is always limited to unity by Fourier relation ($\Delta\omega_{cav} = 1/\tau_{out}$), a value commonly referred to as the 'time-bandwidth limit'[1–3]. This inherent limitation simply dictates that long storage times unavoidably imply narrow input bandwidths, while large bandwidths are unfortunately retained only for short periods of time. Photonics is particularly affected by the time-bandwidth limit. On the one hand, long interaction times are required for storage of optical pulses and efficient light-matter interaction (such as absorption, emission and nonlinear optical effects). On the other hand, broadband signals are desirable since they are normally associated with larger amount of information.

Over the last twenty years, several designs aiming at overcoming this limitation have been investigated. One approach consists of leveraging slow-light waveguides. Such systems exploit the characteristic refractive index dispersion near resonances, due to intrinsic electronic transitions[4–6] or induced by stimulated Brillouin[7,8] or Raman scattering[9,10], or Bragg reflections in periodic structures[11], to slow down the propagation speed of light (namely its group velocity) in the medium. All of these systems operate in the 'waveguide regime', even when based on coupled resonator waveguides, where there is single-pass light propagation and continuous dispersion. In this regime, the time-bandwidth performance of the device is inherently different from that of isolated resonators. Rather than being coupled to a resonant mode, light undergoes a *delay* that can be extended by either increasing group index or propagation length. Nevertheless, these systems are still characterized in terms of a group-index–bandwidth limit[12] or a time-delay–bandwidth–footprint limit. In both of these terms, slow-light waveguides are intrinsically limited, and – similarly to resonant systems – the achievable delay times remain inversely proportional to the waveguide's bandwidth, $\Delta t \sim \Delta\omega^{-\alpha}$, where typically $\alpha = 2$ or $3^{4,13,14}$. Here, the trade-off arises from pulse temporal broadening owing to various dispersion phenomena (2nd and 3rd order dispersion, dispersion of gain/absorption), preventing significant slowing-down of broadband signals[4,7,13,14].

An attempt to overcome the time-bandwidth limit was reported some time ago[15]. That scheme made use of temporal *adiabatic* switching of a system between two reciprocal states: a large-bandwidth–short-storage-time state (low quality factor, $Q$, state) and a narrow-bandwidth–long-storage-time one (high-$Q$ state). However, while the time-bandwidth limit was marginally exceeded (by a factor of 2 or less), the spectral and temporal shapes of the released pulse were not preserved; rather, they strongly depended on the property of the reopened cavity, leading to substantial distortions of the released pulse[15–20]. Crucially, the simultaneous storage of *multiple* pulses in the system cannot be achieved with that scheme: while the bandwidth of the first pulse is adiabatically compressed, a second pulse cannot be injected into the device.

More recently, a proposal[21] for arbitrarily overcoming the time-bandwidth limit of resonant systems was put forward based on breaking *Lorentz reciprocity*[22,23], without accompanying adiabaticity or signal distortion limitations. This theoretical proposal has reinvigorated a debate whether (or not) the time-bandwidth limit can be exceeded in resonant systems[24,25]. However, much of this recent theoretical activity on nonreciprocal resonators focused on time-invariant systems.

In this work, we provide the first experimental confirmation that inducing nonreciprocity by breaking the time-invariance in a cavity system can overcome the 'fundamental' time-bandwidth limit. Using a macroscopic, fiber-optic resonator, in which Lorentz reciprocity is broken by suitable time modulation (i.e. time variant system), we report a TBP above the fundamental limit of ordinary reciprocal cavities by a factor of 30, solely limited by current experimental constraints of our setup. The *non-adiabatic* switching from fully open to fully closed state does not affect the spectral and temporal properties of the injected pulses, and allows for simultaneously storing multiple pulses. Overall, our resonant system is Lorentz-nonreciprocal owing to breaking of its time-invariance[26–29], allowing us to decouple cavity photon lifetime from cavity acceptance bandwidth.

A general definition of the TBP can be obtained in terms of the system's loading ($\rho_L$) and decay ($\rho_D$) energy rates as:

$$TBP = \Delta\omega_{\text{acc}} \tau_{\text{out}} = \frac{\Delta\omega_{\text{acc}}}{\Delta\omega_{\text{cav}}} = \frac{\rho_L}{\rho_D}. \tag{1}$$

where $\Delta\omega_{\text{acc}}$ and $\Delta\omega_{\text{cav}}$, are the full width at half maximum of the Lorentzian functions associated, through the Fourier transform, respectively to the loading and decay curves of the intra-cavity energy (see Supplementary Information). As it is well known, the decay of the energy stored within a cavity is caused by the loss of power through radiative (transmission through coupling elements such as mirrors, couplers etc.) and non-radiative processes (absorption losses), which are taken into account by the out-coupling $\rho_{out}$ and intrinsic $\rho_0$ energy decay rates, respectively. The total decay rate can therefore be expressed as: $\rho_D = \rho_{out} + \rho_0$. Analogously, the loading curve depicts how fast the intra-cavity energy would exponentially grow if the resonator was 'fed' through the same processes reversed in time. As a result, loading rate can be expressed as $\rho_L = \rho_{in} + \rho_0$, with $\rho_{in}$ and $\rho_0$ the in-coupling rate and intrinsic loading rate of energy respectively. In fact, even if the incident light is an arbitrary waveform, the optimum coupling in a resonator is the time reversed version of the decay curve, which corresponds to an exponentially increasing waveform[26,30]. Therefore, the acceptance bandwidth we must considered is the FWHM of the Lorentzian function obtained from the Fourier transform of the loading curve. It represents the maximum Lorentzian linewidth allowed at the input by the resonator in one free spectral range (FSR). In ordinary (reciprocal) resonant devices, $\rho_{out} = \rho_{in}$[21,31] and therefore the two curves are identical ($\rho_L = \rho_D$). The system is said time-reversal symmetric and, as a result, $\Delta\omega_{acc} = \Delta\omega_{cav}$ and TBP = 1. For such a system, the bandwidth of an incoming pulse must be equal to or smaller than the measured resonance linewidth in order to be entirely coupled in the reciprocal cavity. However, in a time-variant nonreciprocal system, we can decouple $\rho_{in}$ from $\rho_{out}$, so that the time-reversal symmetry no longer holds, and the loading process can be made faster than the decay process, meaning that $\rho_L > \rho_D$. In this case, the system can show an arbitrary large TBP. This concept is schematically illustrated in Fig. 1a.

We experimentally implemented such a system, at telecommunication wavelengths (1.55 µm), based on a Sagnac interferometer connected to a highly reflective element, also known as Figure-9 fiber cavity[32–35] (see Methods). We use this simple known fiber configuration, similar in some ways to a recirculating fiber loop, as a platform to demonstrate for the first time a corroboration of the theory that a resonant system with a nonreciprocal coupling can exhibit an arbitrarily high TBP[21]. We break the time-invariance by using localized time-varying phase modulation asymmetrically positioned inside the Sagnac loop, thereby inducing nonreciprocity in the overall system (see Supplementary Information). This allows us to change in time the in-coupling/out-coupling energy rate of the resonator, which results in a dynamic control of the cavity $Q$-factor. The induced change is non-adiabatic because the modulation is shorter than the round trip time of the cavity ($T_{RT}$), which is the inverse of the frequency separation between the resonance lines[36,37].

As depicted in Fig. 1b, light pulse incident to the R port of the 50/50 coupler is split in clockwise (CW) and counter-clockwise (CCW) pulses travelling through the loop. If no phase modulation takes place, the two pulses travel the exact same path and constructively interfere at the reflection port R, as such exiting the resonator. However, if the phase modulator is electrically gated to shift by $\pi$ the phase of the CCW pulse only, then the pulses constructively interfere at the transmission port T of the coupler and the whole light pulse is directed to the reflective element. During the modulator gating time, say $t_1 < t < t_2$, the system is thus a *completely open cavity* capable of fully accepting the pulse without any reflection. When the pulse is reflected back into the Sagnac interferometer by the reflective element, if no other gate signal is applied to the modulator, the CW and CCW pulses again interfere constructively at the T port, meaning that the light pulse is trapped (Fig. 1c). The system hereafter, say time $t > t_2$, acts as a *completely closed cavity* formed by the Sagnac interferometer and the reflective element. We can extract the pulses from the resonator after a desired number of cavity round trips (RT) by gating once again the phase modulator (for $t_2 < t < t_3$), leading to switch the constructive interference to the R port, as illustrated in Fig. 1d. It is important to note that during each stage of operation, i.e. injection, storing and release, the system is reciprocal and therefore the acceptance bandwidth coincides with the cavity bandwidth. However, the breaking of time invariance renders the system nonreciprocal[22], since the system exhibits two different bandwidths during the injection and the storing stages. As any other fiber optic resonator based on standard single mode fiber, this system is subject to the limitations dictated by dispersion and nonlinearity. Specifically, in case of storing of a data pattern made of a sequence of ultra-short pulses, the storage time would be limited by dispersion since the pulses would broaden in time, which could cause the loss of information originally contained in the pattern. This can be dealt, to a certain amount, by dispersion management of the cavity. Besides, an excessively high peak power would induce nonlinear effects, leading to spectral broadening and distortion of the optical bit stream[38]. However, this work does not aim at proposing a novel device, rather at demonstrating a theoretical principle according to which a resonant system with a nonreciprocal coupling can exhibit an arbitrarily high TBP. Therefore, in the experiments, we chose the peak power and the pulse duration in order to have negligible effect of dispersion and nonlinearity, although, in the context of the time-bandwidth performance, in theory, there is no restriction regarding the peak power and the pulse duration.

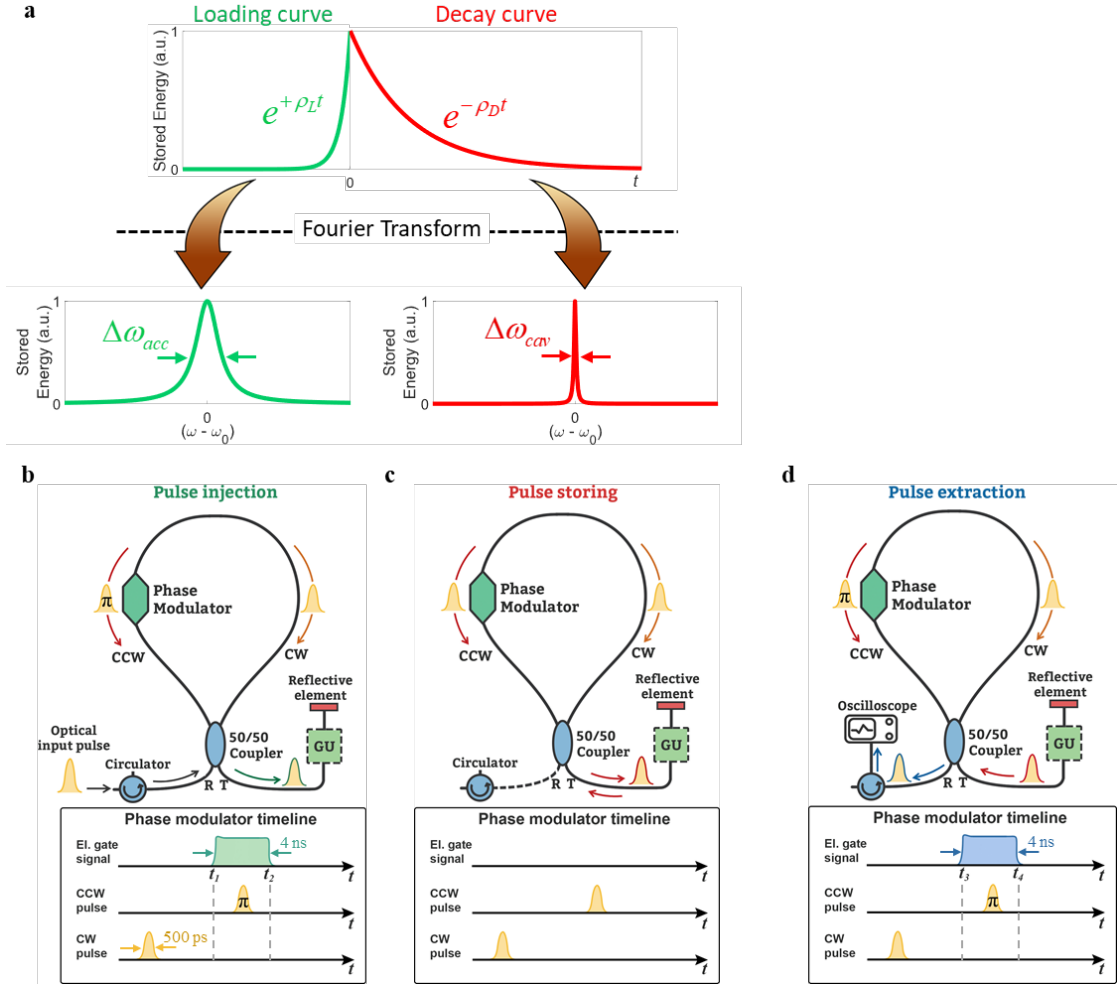

**Fig. 1. Exponential loading and decay curves of a time-variant nonreciprocal resonator and its corresponding implementation in a Figure-9 resonator. a**, If the loading rate is higher than the decay rate, the exponential energy loading process is faster than the decay process, and their associated bandwidths, $\Delta\omega_{acc}$ and $\Delta\omega_{cav}$, respectively, are different, with $\Delta\omega_{acc} > \Delta\omega_{cav}$. **b**, Injection - The optical input pulse is split in two counter-propagating pulses (CW and CCW) in the fiber loop and is fully coupled in the cavity owing to constructive interference at the T port. This is obtained by imparting a π phase shift solely to the CCW pulse using a gated phase modulator. **c**, Storing - Once loaded, if no other gate signal is applied to the modulator, the CW and CCW pulses interfere constructively at the T port and the pulse is stored in the resonator until it is dissipated through internal loss. **d**, Extraction - The pulse is extracted after a desired number of RTs by opening again the cavity, i.e. applying a second "gate" signal to the phase modulator to the CCW portion of the pulse. The gain unit (GU), represented by the green dashed block, is not actually present for the initial, passive experiments (e.g. Fig. 2a), but is incorporated for subsequent experiments to partially compensate for the dissipative loss (e.g. Fig. 2b).

To express the TBP as a function of the parameters that characterize the Figure-9 resonator, it is convenient to define the energy rates in terms of the in- and out-coupling transmission coefficients of the Sagnac interferometer, $\alpha_{in}$ and $\alpha_{out}$, respectively. We have: $\rho_L = \alpha_{in}/T_{RT} + 1/\tau_0$ and $\rho_D = \alpha_{out}/T_{RT} + 1/\tau_0$, (see Supplementary Information), where $\tau_0$ is the internal, non-radiative decay time, usually associated with absorption or energy dissipation inside the cavity. Here $\tau_0$ also takes into account the decay of energy due to the small leakage from the reflective element. When the resonator is in the fully open state at time $t_1 < t < t_2$, we have $\alpha_{in}(t_1) = \alpha_{out}(t_1) = 1$. We can note that the system is actually not a cavity in this case, but an ordinary delay line/waveguide with a reflective termination, and the delay experienced by the pulse is simply $T_{RT}$. It thus seems not possible to associate a linewidth to the cavity in the open state. However, as we have already mentioned, the acceptance bandwidth is by definition the FWHM of the Lorentzian profile associated to the energy loading process of the cavity. In this way, a linewidth related to a "fictitious" loading resonant mode, which is quantified by the in-coupling energy rate ($\rho_{in}$) and the intrinsic energy rate ($\rho_0$), can always be associated to the cavity. This is true even in the extreme case of fully open state where $\alpha_{in}(t_1) = 1$, and, therefore, $\rho_L(t_1) = 1/T_{RT} + 1/\tau_0$. Once the pulse is coupled into the resonator and the system is switched to the fully closed state at time $t_2$, we have $\alpha_{in}(t_2) = \alpha_{out}(t_2) = 0$ and $\rho_D(t_2) = 1/\tau_0$. Thus, for this time-variant system the TBP reduces to the following simple relation:

$$TBP = \frac{\rho_L(t_1)}{\rho_D(t_2)} = \frac{\tau_0}{T_{RT}} + 1 = \frac{F_{closed}}{2\pi} + 1 \qquad (2)$$

with $F_{closed}$ the finesse of the closed cavity. As a result, by decoupling in time the cavity photon lifetime $\tau_{out}$ (or equivalently the cavity bandwidth $\Delta\omega_{cav}$) from the cavity acceptance bandwidth $\Delta\omega_{acc}$, such that $\rho_L(t_1) > \rho_D(t_2)$, the TBP of the system can be higher than 1. We stress that, even if the actual bandwidth we can physically couple inside the cavity is in practice only limited by the operating frequency region of the 50/50 coupler, the acceptance bandwidth that has to be considered in calculating the TBP is the FWHM of the Lorentzian profile associated to the energy loading process. It is thus not given by the bandwidth of the incoming pulse.

The experimental setup is described in details in the methods. The input to the resonator consists of 500 ps Gaussian pulses. Since according to eq. (2), the cavity finesse limits the TBP, we experimentally control $F_{closed}$ by inserting an optical amplifier (EDFA) inside the resonator. As such, we can tune $\tau_0$ by varying the EDFA gain. We measure the cavity RT time to be 48 ns and 120.3 ns, without and with the EDFA respectively. It is important to note that the addition of an EDFA is a mean to overcome relatively high absorption losses, adding gain without exceeding the losses, while not affecting the general principle. In fact, an analogous amplification would never increase the TBP beyond one in a reciprocal resonator, as more power would simply also leak out the system at every round trip.

We assess the performance of the system by measuring the energy of the pulse released after different numbers of RTs. Figure 2a shows the result for the passive cavity (no EDFA). The exponential decay fit of the experimental data corresponds to a decay time $\tau_0$ of about 65.69 ns, which allowed us to extract a pulse above the noise level after up to 10 RTs. This corresponds to a closed cavity decay-time of about 1.37 times longer than the cavity RT time, leading to a TBP of 2.37. According to Eq. (2), the maximum achievable TBP can be in principle infinite, providing an infinitely long closed-cavity decay time $\tau_0$, i.e. a loss-less cavity. However, in our case $\tau_0$ is limited by a technological constraint, specifically the absorption losses at the modulator measured to be ~3.17 dB/RT. We therefore use the active cavity configuration (with EDFA) to support the claim of arbitrarily large TBP by experimentally controlling the decay time of the system. We progressively adjust the power of the EDFA to partially compensate the intra-cavity loss over three different steps resulting in a net loss of 0.4, 0.25, 0.15 dB/RT. The measurements are shown in Fig. 2b, where the experimental data is normalized to the energy of the pulse extracted after the first cavity RT. As the addition of the EDFA increases $T_{RT}$, according to Eq. (2), this might actually reduce the TBP of the system. However, the significant increase in $\tau_0$ allows sustaining the pulse for up to 120 RTs (red curve). The decay time strongly increases from 65.69 ns up to 3.57 µs, resulting in a maximum TBP of 30.7. For this measurement, the period of the input pulse train lies between 30 and 31 RTs, to avoid time overlap between the intra-cavity pulse in its 31$^{st}$ round trip and the new incoming input pulse. In this way, we can couple multiple pulses in the resonator and extract an individual pulse after more than 30 RTs without affecting the others.

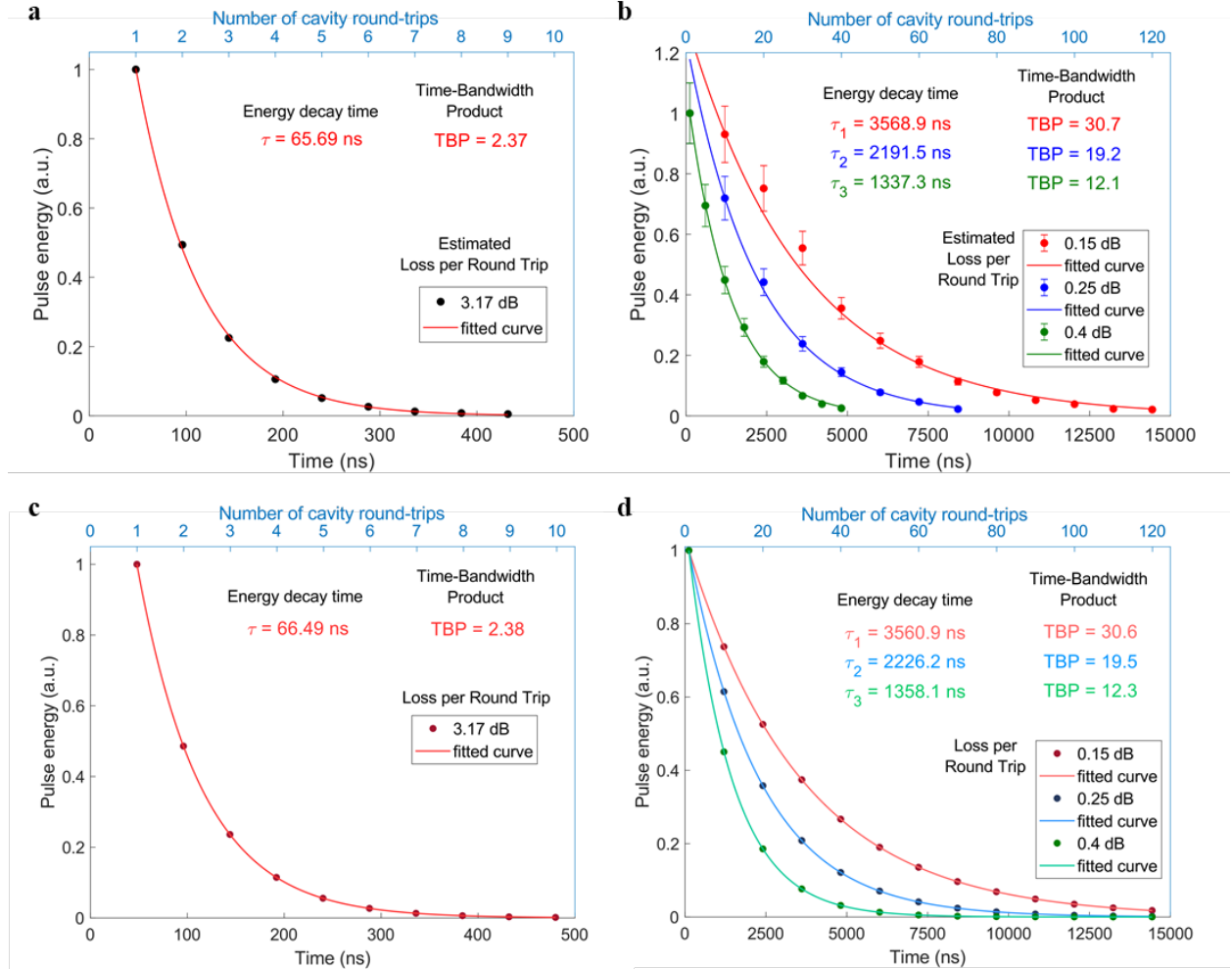

**Fig. 2. Energy decay curves related to the 500 ps Gaussian pulse. a**, Experimentally measured decay curve for the pulses extracted from the full polarization maintaining fiber *passive* resonator at every round trip time ($T_{RT}$ = 48 ns). **b**, Experimentally measured decay curves for the pulses extracted from the amplified resonator every ten round trips (with $T_{RT}$ = 120.3 ns) for different values of loss per RT. Error bars in **b** come from fast polarization rotation due to the non-polarization maintaining erbium doped fiber in the gain unit, resulting in a 20% uncertainty. **c**, Simulated decay curve of the passive cavity with the same actual value of loss/RT as for **a**. **d**, Simulated decay curves of a passive cavity configuration with the same actual value of loss/RT as for **b**.

In principle, we could achieve an even higher TBP value by intensifying the pump power of the EDFA as to fully compensate the round-trip loss. Under these conditions the TBP is higher, but now limited by dispersion, nonlinear effects and the amplification of noise by the EDFA. However, in practice, we were limited by the gain saturation of the doped fiber. This effect can be seen in Fig. 2b for the configuration with 0.15 and 0.25 dB/RT of effective losses. In fact, here the pulses retrieved at the first RT have energies sufficiently high to saturate the gain of the amplifier, which cannot compensate the cavity losses in the same way as for the pulses extracted after more RTs. This results in higher effective cavity losses at the first experimental point, which we therefore excluded from the fit. Further increasing the diode pump power would have affected even more points, misleading the estimate of the TBP.

In order to confirm this concept, we conducted detailed simulations of the pulse storing operation using VPIphotonics software (see Methods for details). Our experimental resonator was numerically modelled in 4 passive configurations (without EDFA): in the first one we have reproduced the exact passive experimental cavity (Fig. 2c), while in the other three configurations we have set the total loss and $T_{RT}$ as to mimic the three values of the experimental active setup (Fig. 2d). For all, the value of the TBP is in excellent agreement with the one calculated after fitting of the experimental data. In particular for Figure 2d, the exponential decay fits almost perfectly the experiments, showing decay times from 1.36 to 3.56 µs as the dissipative losses progressively decrease. The simulation not only confirm the improvement in TBP but also that we can indeed treat our active cavity as a passive cavity with reduced dissipative losses.

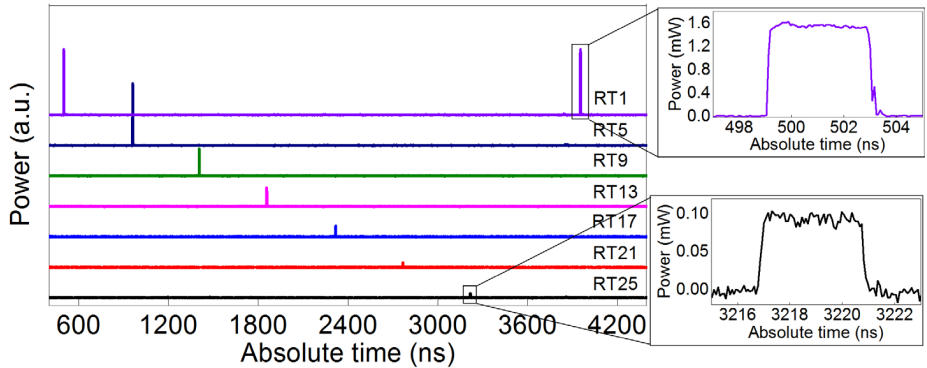

**Fig. 3. Temporal traces over one period of the optical pulse train extracted after different RTs.** The diode pump power of the gain unit was adjusted to obtain a configuration with about 0.5 dB of loss per round trip.

In Fig. 3 we provide an example showing the temporal traces of a 4 ns squared pulse stored in the resonator and extracted after different RTs, with loss of about 0.5 dB/RT. The pulse can be extracted after up to 25 RTs and no leakage is observed between two subsequent extracted pulses. This confirms that we can couple the entire pulse energy ($\alpha_{in} \approx 1$) without any out-coupling loss ($\alpha_{out} \approx 0$), switching the cavity from the completely open to the completely closed state. For this specific measurement we used a longer and square-shaped pulse because the acquisition memory of our oscilloscope was not sufficient to detect the 500 ps long Gaussian pulses over the entire time period of the pulse train (about 3.6 μs).

Fundamentally different from time-variant devices based on adiabatic tuning[15–20], here we do not need to adiabatically compress the input pulse bandwidth to match the closed cavity resonance and avoid scattering between different resonant modes. Indeed, we are in a *non-adiabatic* regime, as $T_{RT}$ is longer than the tuning time, which is given by the rising time of the phase modulator. Moreover, with $T_{RT}$ being longer than the pulse duration, the injected pulse does not interfere with itself and cannot 'see' the closed-cavity resonant modes. Therefore, the pulse does not need to adapt to the closed-cavity resonances and, once released, it exhibits a spectrum that is unaffected by the switching between the two different cavity states. To clearly show that the characteristics of the released pulses are preserved over all the RTs, we collected temporal waveforms and radio-frequency (RF) spectra (see Methods for details on the measurement technique) of the 500 ps Gaussian pulse after 1, 40 and 80 RTs (Figs. 4b, 4c and 4d) and plotted together with those of the pulse collected before entering the cavity (Fig. 4a). The product of the pulse duration and bandwidth (FWHM) retrieved from the Gaussian fit was always about 0.44 for the investigated RTs, confirming that the pulse does not suffer any measurable distortions.

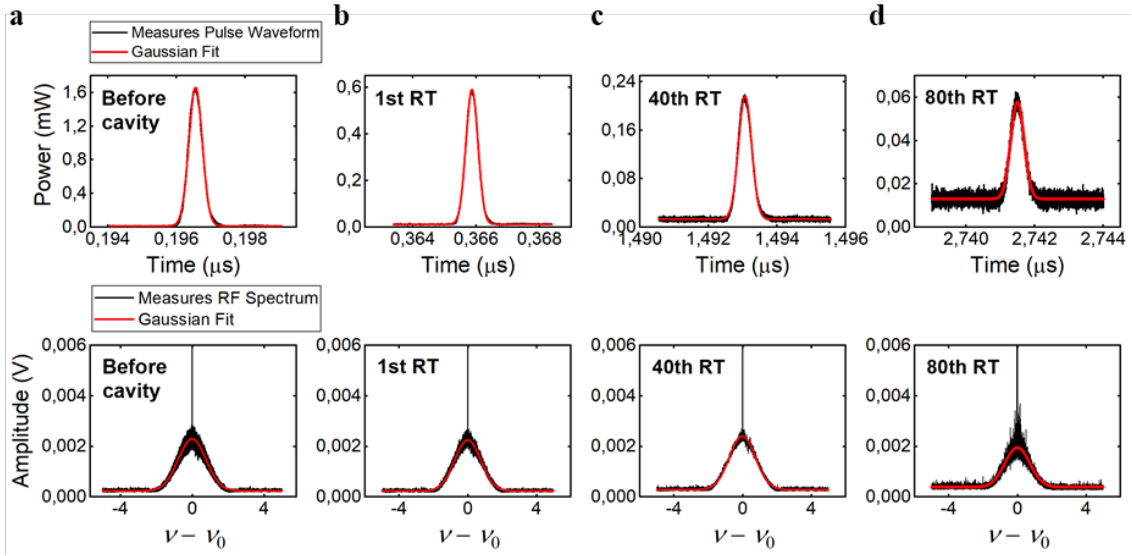

**Fig. 4. Pulse waveforms and radio-frequency spectra**. Traces acquired before the cavity (**a**) and after 1 RT (**b**), 40 RTs (**c**) and 80 RTs (**d**). The product of the pulse duration and the bandwidth (FWHM) gives values close to the transform-limited pulse for all the three cases.

In conclusion, we experimentally demonstrated that breaking the time-invariance in a resonant system, thus inducing nonreciprocity, allows to arbitrarily overcome the time-bandwidth limit[21] by completely decoupling the input energy rate from the cavity storage time. We used localized time-varying phase modulation to dynamically control the *Q*-factor of a macroscopic fiber resonator, which we switched from a completely open to a completely closed state. We proved that the value of the TBP of an individual resonator is ultimately equal to $F_{closed}/2\pi + 1$ and can be increased at will above the limit, provided that internal, dissipative losses are kept sufficiently low. Mitigating for these dissipative losses with a gain unit, we reported a TBP thirty (30) times above the 'fundamental' time-bandwidth limit of ordinary resonators – limited only by current experimental constraints of our setup. Additionally we could simultaneously store and manipulate *multiple* pulses, a key capability missing from previous adiabatic cavity modulation schemes[15,20]. When retrieved, the pulses did not exhibit detectable temporal and spectral distortions. The presented scheme may thus open the path for a wealth of applications – both fundamental and applied, throughout physics and engineering – where large bandwidths, ultrafast response times, long storage durations, high sensitivities and strong wave-matter interactions are simultaneously desired[39].

**References:**

1. Quimby, R. S. *Photonics and Lasers: An Introduction* (John Wiley & Sons, 2006), chap. 16.
2. Van, V. *Optical microring resonators: theory, techniques, and applications*. *Contemporary Physics* **59**, (CRC Press, 2017), chap. 2.
3. Demtröder, W. *Laser Spectroscopy 1*. (Springer, 2014), chap. 3.
4. Khurgin, J. B. Slow light in various media: a tutorial. *Adv. Opt. Photonics* **2**, 287–318 (2010).
5. Bigelow, M. S., Lepeshkin, N. N. & Boyd, R. W. Superluminal and slow light propagation in a room-temperature solid. *Science* **301**, 200–202 (2003).
6. Camacho, R. M., Pack, M. V & Howell, J. C. Slow light with large fractional delays by spectral hole-burning in rubidium vapor. *Phys. Rev. A* **74**, 033801 (2006).
7. Okawachi, Y. *et al.* Tunable all-optical delays via brillouin slow light in an optical fiber. *Phys. Rev. Lett.* **94**, 153902 (2005).

8. González-Herráez, M., Song, K.-Y. & Thévenaz, L. Optically controlled slow and fast light in optical fibers using stimulated Brillouin scattering. *Appl. Phys. Lett* **87**, 081113 (2005).
9. Sharping, J. E., Okawachi, Y. & Gaeta, A. L. Wide bandwidth slow light using a Raman fiber amplifier. *Opt. Express* **13**, 6092–6098 (2005).
10. Turukhin, A. V *et al.* Observation of ultraslow and stored light pulses in a solid. *Phys. Rev. Lett.* **88**, 023602 (2002).
11. Baba, T. Slow light in photonic crystals. *Nat. Photonics* **2**, 465–473 (2008).
12. Schulz, S. A. *et al.* Dispersion engineered slow light in photonic crystals: a comparison. *J. Opt.* **12**, 104004 (2010).
13. Boyd, R. W., Gauthier, D. J., Gaeta, A. L. & Willner, A. E. Maximum time delay achievable on propagation through a slow-light medium. *Phys. Rev. A* **71**, 023801 (2005).
14. Khurgin, J. B. Performance limits of delay lines based on optical amplifiers. *Opt. Lett.* **31**, 948–950 (2006).
15. Xu, Q., Dong, P. & Lipson, M. Breaking the delay-bandwidth limit in a photonic structure. *Nat. Phys.* **3**, 406–410 (2007).
16. Tanaka, Y. *et al.* Dynamic control of the Q factor in a photonic crystal nanocavity. *Nat. Mater.* **6**, 862–865 (2007).
17. Upham, J. *et al.* The capture, hold and forward release of an optical pulse from a dynamic photonic crystal nanocavity. *Opt. Express* **21**, 3809–3817 (2013).
18. Yanik, M. F. & Fan, S. Stopping Light All Optically. *Phys. Rev. Lett.* **92**, 083901 (2004).
19. Yanik, M. F. & Fan, S. Stopping and storing light coherently. *Phys. Rev. A* **71**, 013803 (2005).
20. Tanabe, T., Notomi, M., Taniyama, H. & Kuramochi, E. Dynamic release of trapped light from an ultrahigh-Q nanocavity via adiabatic frequency tuning. *Phys. Rev. Lett.* **102**, 043907 (2009).
21. Tsakmakidis, K. L. *et al.* Breaking Lorentz reciprocity to overcome the time-bandwidth limit in physics and engineering. *Science* **356**, 1260–1264 (2017).
22. Caloz, C. *et al.* Electromagnetic Nonreciprocity. *Phys. Rev. Appl.* **10**, 047001 (2018).
23. Potton, R. J. Reciprocity in optics. *Reports Prog. Phys.* **67**, 717–754 (2004).
24. Mann, S. A., Sounas, D. L. & Alù, A. Nonreciprocal Cavities and the Time-Bandwidth Limit. *Optica* **6**, 104–110 (2019).

25. Tsang, M. Quantum limits on the time-bandwidth product of an optical resonator. *Opt. Lett.* **43**, 150–153 (2017).

26. Leuchs, G. & Sondermann, M. Time-reversal symmetry in optics. *Phys. Scr.* **85**, 058101 (2012).

27. Jalas, D. *et al.* What is-and what is not-an optical isolator. *Nature Photonics* **7**, 579–582 (2013).

28. El-Ganainy, R. *et al.* Non-Hermitian physics and PT symmetry. *Nat. Phys.* **14**, 11–19 (2018).

29. Bahari, B. *et al.* Nonreciprocal lasing in topological cavities of arbitrary geometries. *Science* **358**, 636–640 (2017).

30. Heugel, S., Villar, A. S., Sondermann, M., Peschel, U. & Leuchs, G. On the analogy between a single atom and an optical resonator. *Laser Phys.* **20**, 100–106 (2010).

31. Zhao, Z., Guo, C. & Fan, S. Connection of temporal coupled-mode-theory formalisms for a resonant optical system and its time-reversal conjugate. *Phys. Rev. A* **99**, 33839 (2019).

32. Cowle, G., Payne, D. & Reid, D. Single-frequency travelling-wave erbium-doped fibre loop laser. *Electron. Lett.* **27**, 229–230 (1991).

33. Cardea, I., Kharitonov, S. & Brès, C.-S. Experimental and theoretical investigation of the operating principles of the Figure-9 laser. in *Advanced Photonics Congress 2018* SoW2H.6 (OSA, 2018). doi:doi.org/10.1364/SOF.2018.SoW2H.6

34. Krzempek, K., Sotor, J. & Abramski, K. Compact all-fiber figure-9 dissipative soliton resonance mode-locked double-clad Er:Yb laser. *Opt. Lett.* **41**, 4995–4998 (2016).

35. Kharitonov, S. & Brès, C.-S. Unidirectional all-fiber thulium-doped laser based on theta cavity and fiber Bragg grating as filtering element. in *Lasers Congress 2016 (ASSL, LSC, LAC)* AM5A.5 (2016). doi:10.1364/ASSL.2016.AM5A.5

36. Gaburro, Z. *et al.* Photon energy lifter. *Opt. Express* **14**, 7270–7278 (2006).

37. Galindo, A. & Pascual, P. *Quantum Mechanics II.* (Springer-Verlag Berlin Heidelberg, 1991), chap. 11.

38. Agrawal, G. P. *Lightwave Technology: Telecommunication Systems. Lightwave Technology: Telecommunication Systems* (2005), chaps. 3, 4 and 7.

39. Tsakmakidis, K. L., Hess, O., Boyd, R. W. & Zhang, X. Ultraslow waves on the nanoscale. *Science* **358**, eaan5196 (2017).

40. Forrester, A. T. Photoelectric Mixing As a Spectroscopic Tool. *J. Opt. Soc. Am.* **51**, 253–259

(1961).

**Author contributions**: Preliminary discussions between K.L.T, H.A, I.C., and D.G. initiated the project. I.C. and D.G. conceived the experiment and developed the theory, with inputs from K.L.T., S.A.S. and J.U.; I.C. performed the experiments, with input from D.G. and S.J.F.; I.C., D.G. and C.S.B. wrote the manuscript. All the authors reviewed and edited the manuscript. C.S.B. supervised the entire project.

**Competing interests:** The authors declare no competing interests.

**Data and materials availability:** The data that support the findings of this study are available from the corresponding authors on reasonable request.

**Methods**

**Description of the experimental setup.** The resonator used in the experiments is made of polarization maintaining fibers with a $T_{RT}$ of about 48 ns, while the reflective element is a fiber Bragg grating with a center wavelength at 1551.3 nm and a stop bandwidth of about 28.2 GHz. The gain unit is made of 90 cm-long Erbium-doped fiber connected with two fused fiber wavelength division multiplexers and pumped by a semiconductor laser diode at 980 nm. The optical input pulse train at 1551.3 nm is obtained from a laser, modulated in intensity to give 500 ps Gaussian pulses with 894 MHz bandwidth, and a repetition rate corresponding to about 30 cavity RTs. We synchronized an electrical pulse of 4 ns with the optical signal to activate the phase modulator when it is traversed by the CCW pulse only. Once extracted, the pulses are detected at the third port of a circulator, placed before the R port of the Sagnac interferometer, by using a high-speed sampling oscilloscope. Both the electrical signals used to drive the phase and the intensity modulator were generated by the same arbitrary waveform generator (AWG) (Tektronix model 7122B). The phase modulator used for the experiments was a $LiNbO_3$ electro-optic modulator (Photline model MPZ-LN-10) with an electro-optic bandwidth of 12 GHz. The synchronization between the electrical "gate" and the optical signal was performed directly from the AWG by imposing a delay on the electrical signal that drove the phase modulator.

**Methods for the measurements of the pulse waveforms and spectra.** The temporal traces were registered by detecting the extracted pulses on a sampling oscilloscope with 20 GHz of optical bandwidth. Given the limited resolution of our OSA, a direct measurement of the pulse spectrum in the optical domain did not provide the suitable resolution to detect variations in the spectrum of the order of the cavity free-spectral range (about 8 MHz). We thus implemented a zero-delay self-heterodyne technique[40], to map the optical spectrum of the pulses into the radio-frequency domain. The pulses retrieved from the resonator were modulated using a 40 GHz Mach-Zehnder modulator to create sidebands at 16 GHz from the central pulse frequency and sent to an Electrical Spectrum Analyzer (ESA). The bottom row of Fig. 4 reports the radio-frequency spectra, given by the convolution of the beating lines acquired with the ESA and centered at the modulation frequency.

**Methods for the numerical simulations.** The simulations were performed using the tool VPItransmissionMaker™ Optical Systems of the software VPIphotonics Design Suite™ whose numerical solver is based on a full-wave analysis. We reproduced the setup in the graphical environment using built-in blocks with customized parameters.

# Supplementary Information

# A nonreciprocal optical resonator with broken time-invariance for arbitrarily high time-bandwidth performance

Ivan Cardea[1], Davide Grassani[1,2], Simon J. Fabbri[1], Jeremy Upham[3], Robert W. Boyd[3], Hatice Altug[4], Sebastian A. Schulz[5], Kosmas L. Tsakmakidis[6], Camille-Sophie Brès[1*].

[1]Ecole Polytechnique Fédérale de Lausanne (EPFL), Photonic Systems Laboratory (PHOSL), Lausanne CH-1015, Switzerland.

[2]Currently with Dipartimento di Fisica, Università degli studi di Pavia, via Bassi 6, 27100 Pavia, Italy.

[3]Department of Physics, University of Ottawa, Ottawa, ON, Canada.

[4]Ecole Polytechnique Fédérale de Lausanne (EPFL), Bionanophotonics System Laboratory (BIOS), Lausanne CH-1015, Switzerland.

[5]School of Physics and Astronomy, SUPA, University of St Andrews, St Andrews, KY169SS, UK.

[6]Solid State Physics section, Department of Physics, National and Kapodistrian University of Athens, Panepistimioupolis, GR - 157 84, Athens, Greece.

*Correspondence to: camille.bres@epfl.ch.

## Table of Contents

**Acceptance bandwidth and time-bandwidth product (TBP) in reciprocal and non-reciprocal time-variant resonant systems**

The time-bandwidth product (TBP) of a resonator, or cavity, is more generally defined as the product between its acceptance bandwidth ($\Delta\omega_{acc}$) and its characteristic decay time ($\tau_{out}$). This very general definition keeps the concept of the bandwidth that the resonator can "accept" separated from what is considered as the cavity bandwidth ($\Delta\omega_{cav}$), usually defined as the inverse of $\tau_{out}$. In fact, as we will show here, these two quantities can be different and completely decoupled. This becomes clear considering the temporal evolution of the energy stored in the cavity and comparing it with the same process reversed in time, to which hereafter we will refer as *loading* process. As illustrated in Figure S1, the energy stored in a resonator at $t = 0$ is $|a_D(0)|^2$, and decays exponentially as $e^{-(\rho D)t}$ (red curve), where $\rho_D = \rho_{out} + \rho_0$ is the energy decay rate due both to non-radiative and radiative processes, which are related to internal absorption and out-coupling from the cavity respectively. The Fourier transform of such exponential decay curve is a Lorentzian function, whose FWHM is just $\rho_D$, which is commonly identified as the cavity bandwidth (i.e. $\Delta\omega_{cav} = \rho_D$). Consider the energy inside the resonator at the time $t = T$, that is $|a_D(T)|^2$, as the final state of the decay process. Then, let's apply the time-reversal operation, which mathematically corresponds to reversing the sign of the time variable in the decay process[1–3]. Doing so, $|a_D(T)|^2$ becomes the initial state of the loading process at the time $t = -T$, that is $|a_D(T)|^2 = |a_L(-T)|^2$. Now, the energy flows and builds up in the resonator as $e^{+(\rho_L)t}$, where $\rho_L = \rho_{in} + \rho_0$ is the energy *loading rate*. If $\rho_L$ is equal to $\rho_D$ the time-reversal symmetry holds (orange-dashed curve in Fig S1). However, if they are different, owing to an asymmetry of the radiative process ($\rho_{in} \neq \rho_{out}$), the time-reversal symmetry is broken leading to nonreciprocity in the system. Particularly, if $\rho_{in} > \rho_{out}$, after a time $T$, the energy in the resonator would be higher than the initial value of the energy in the decay process ($|a_L(0)|^2 > |a_D(0)|^2$ – green-dashed curve in Figure S1). It should be noticed that, this latter scenario require the breaking of one of the two hypothesis on which reciprocal resonant systems are based, i.e. linearity or time-invariance[1,4].

In the same way as in the energy decaying process, through the Fourier transform, we can associate a bandwidth to this *loading* process given by the FWHM of the corresponding Lorentzian function, which in this case is equal to $\rho_L$, and represents the acceptance bandwidth of the resonator. In fact, being $a_L(t)$ the field inside the resonator, we can write:

$$a_L(t) = a_L(0)e^{\frac{\rho_L}{2}t}e^{j\omega_0 t}u(-t) \tag{S1}$$

where $u(t)$ is the Heaviside step function and $\omega_0$ is the resonant frequency. The corresponding Fourier transform is:

$$a_L(\omega) = \int_{-\infty}^{\infty} a_L(t)e^{-j\omega t}dt = \int_{-\infty}^{0} a_L(0)e^{\left[\frac{\rho_L}{2}+j(\omega_0-\omega)\right]t}dt \Rightarrow$$

$$a_L(t) = a_L(0)\left[\frac{e^{\left[\frac{\rho_L}{2}+j(\omega_0-\omega)\right]t}}{\frac{\rho_L}{2}+j(\omega_0-\omega)}\right]_{-\infty}^{0} = -\frac{a_L(0)}{j(\omega-\omega_0)-\frac{\rho_L}{2}}. \tag{S2}$$

The expression of the energy inside the cavity is obtained by taking the square modulus of $a_L(t)$:

$$W = |a_L(t)|^2 = \frac{|a_L(0)|^2}{(\omega-\omega_0)^2 + \frac{\rho_L^2}{4}}. \tag{S3}$$

The loading process is thus characterized by a Lorentzian function too, and its FWHM represents the acceptance bandwidth of the resonator:

$$\Delta\omega_{acc} = \rho_L \tag{S4}$$

In reciprocal resonant systems, $\rho_{in} = \rho_{out}$, therefore $\rho_L = \rho_D$ and $\Delta\omega_{acc} = \Delta\omega_{cav}$. The TBP of such reciprocal system ($TBP_R$) is equal to 1:

$$TBP_R = \Delta\omega_{acc}\tau_{out} = \frac{\Delta\omega_{acc}}{\Delta\omega_{cav}} = 1 \tag{S5}$$

However, if $\rho_{in} \neq \rho_{out}$, the resonant system is nonreciprocal and, if $\rho_{in} > \rho_{out}$, the loading process reaches a final state at $t = 0$ that is higher than the initial state of the decaying process (see Fig. S1). Therefore, the acceptance bandwidth results to be larger than the cavity bandwidth, leading to a TBP for such nonreciprocal system ($TBP_{NR}$) given by:

$$TBP_{NR} = \frac{\Delta\omega_{acc}}{\Delta\omega_{cav}} = \frac{\rho_{in}+\rho_0}{\rho_{out}+\rho_0} > 1 \tag{S6}$$

In the Figure-9 resonator the nonreciprocity is obtained by breaking the time-invariance of the system. As explained in Figure S2a, either in the open and the closed state the resonator exhibits identical in- and out-coupling energy rate ($\rho_{\text{in-open}} = \rho_{\text{out-open}}$ and $\rho_{\text{in-closed}} = \rho_{\text{out-closed}}$). However, $\rho_{\text{in-open}} > \rho_{\text{out-closed}}$. As a result, the resonator can be filled with energy at a rate $\rho_{L\text{-open}}$, but once the system is switched to the closed state, the energy decay occurs at a rate $\rho_{D\text{-closed}} < \rho_{L\text{-open}}$ as illustrated in Figure S2b. Therefore, although the resonant system is reciprocal at any given time since its time-reversal symmetry is never broken, it can be considered nonreciprocal given the temporal switching between the open and closed states. This nonreciprocity induced by the broken time-invariance of the system leads to a decoupling of the acceptance bandwidth from the characteristic decay rate of the cavity, allowing to couple energy in the system at a rate higher than the value imposed by the cavity bandwidth[5]. We stress that, the exponentially increasing loading curve represents the optimum coupling of a resonator, not the actual coupling of energy within the cavity, which can take any form in time domain and occur only through radiative processes, since the internal (non-radiative) absorption is an irreversible process.

**Transmission and reflection coefficients of the Figure-9 resonator**

In general, the Figure-9 cavity is formed by a Sagnac interferometer, or fiber loop mirror, connected to a highly reflecting element (see left part of Figure S3). For the sake of simplicity, the structure of the Figure-9 cavity can be seen as a Fabry-Pérot cavity, as shown in Figure S3, in which the fiber loop mirror represents one of the two reflecting elements with reflection, transmission and attenuation given by those of the Sagnac interferometer. The fiber optic system used in the experiments was based on such a cavity using a high reflectivity fiber Bragg grating (FBG) connected to one of the input ports of a fiber loop mirror (port T in Figure S3). The bandwidth of the FBG was chosen to be large enough to cover the pulse bandwidth, but sufficiently narrow to filter out the amplified spontaneous emission generated by the doped fiber, which, otherwise, would have overwhelmed the signal after few cavity round trips, dramatically reducing the signal-to-noise ratio.

The reflection and transmission coefficients of the fiber loop mirror can be found considering a monochromatic light wave ($E_{\text{in}}$) of frequency $\omega$ incident on the port R of the directional coupler, as schematically illustrated in Figure S3. The amplitude of the wave is split by the coupler in two parts that travel through the same physical path in opposite direction to recombine again in the coupler. Assuming that there is no relative phase delay between the two counter-propagating fields and, recalling that light coupled across the coupler suffer a $\pi/2$ phase lag with respect to light travelling straight through it, we can write the complex amplitudes of the reflected and transmitted fields as following:

$$A_R = j\kappa\tau\sqrt{a_p}\,A_{in}e^{-j\phi_p} + j\kappa\tau\sqrt{a_p}\,A_{in}e^{-j\phi_p} = 2j\kappa\tau\sqrt{a_p}\,A_{in}e^{-j\phi_p} \tag{S7}$$

$$A_T = \tau^2\sqrt{a_p}\,A_{in}e^{-j\phi_p} - \kappa^2\sqrt{a_p}\,A_{in}e^{-j\phi_p} = (\tau^2 - \kappa^2)\sqrt{a_p}\,A_{in}e^{-j\phi_p} \tag{S8}$$

where $\kappa$ and $\tau$ are the cross and straight-through field coupling coefficient of the coupler. The losses in the Sagnac interferometer ($\alpha$) give the attenuation factor ($a_p = e^{-\alpha L}$), while $\phi_p$ is the phase delay experienced by the fields during the propagation between the ports 1 and 2 of the coupler. The *field* reflection and transmission coefficients of the fiber loop are easily found from the above equations:

$$\frac{A_R}{A_{in}} = r_{FL} = j2\kappa\tau\sqrt{a_p}\,e^{-j\phi_p}; \qquad \frac{A_T}{A_{in}} = t_{FL} = (\tau^2 - \kappa^2)\sqrt{a_p}\,e^{-j\phi_p}\ . \tag{S9}$$

The corresponding *power* reflection and transmission coefficients are obtained by taking the square modulus of $r_{FL}$ and $t_{FL}$:

$$|r_{FL}|^2 = 4\kappa^2\tau^2 a_p; \qquad |t_{FL}|^2 = (\tau^2 - \kappa^2)^2 a_p \tag{S10}$$

for which the following relation holds: $|t_{FL}|^2 + |r_{FL}|^2 = a_p$.

In our experiment, we used a 50/50 coupler, that is $\kappa^2 = \tau^2 = 0.5$. Inserting these values in the equation (S10), in absence of any phase difference between the two waves in the Sagnac interferometer, the incident field is totally reflected due to the constructive interference at the R port of the coupler, while the transmission at the T port of the fiber loop mirror is zero ($|r_{FL}|^2 = a_p$ and $|t_{FL}|^2 = 0$). In this configuration, the cavity can be considered completely "closed", because the field incident upon it is totally reflected, apart from an attenuation factor.

If there is a relative phase difference equal to $\pi$ between the two counter-propagating fields, from equations (S7) and (S8) we get:

$$A_R = j\kappa\tau\sqrt{a_p}\,A_{in}e^{-j\phi_p} + j\kappa\tau\sqrt{a_p}\,A_{in}e^{-j\phi_p}e^{j\pi} = 0 \tag{S11}$$

$$A_{\mathrm{T}} = \tau^2 \sqrt{a_{\mathrm{p}}} A_{\mathrm{in}} e^{-j\phi_{\mathrm{p}}} - \kappa^2 \sqrt{a_{\mathrm{p}}} A_{\mathrm{in}} e^{-j\phi_{\mathrm{p}}} e^{j\pi} = (\tau^2 + \kappa^2)\sqrt{a_{\mathrm{p}}} A_{\mathrm{in}} e^{-j\phi_{\mathrm{p}}} \tag{S12}$$

and assuming an ideal coupler ($\tau^2 + \kappa^2 = 1$), we obtain:

$$|r_{\mathrm{FL}}|^2 = 0; \qquad |t_{\mathrm{FL}}|^2 = a_{\mathrm{p}}. \tag{S13}$$

Thus, the incident field is totally transmitted (and partially attenuated) through the fiber loop mirror due to the constructive interference at the port T of the coupler (regardless the value of the coupling coefficient). In this state the cavity is then completely "open".

**Energy rate coefficients of the time-variant Figure-9 resonator**

In the reported experiments, the localized phase variation was provided by an electrically-driven phase modulator, placed at an offset position from the loop midpoint to ensure that the phase shift was imparted only to one of the two counter-propagating pulses. In this way we were able to change in time the reflection and transmission coefficients of the fiber loop, similarly to the mechanism used to switch optical pulses in Terahertz optical asymmetric demultiplexers (TOADs)[6]. In the analogy previously used, the fiber loop represents the front mirror of the Fabry-Pérot resonator, and we use the time-variant phase modulation to dynamically control the Q-factor of the cavity. In our system, the pulse duration is smaller than the cavity round trip time and we can entirely inject the incoming pulse within the cavity, by driving the modulator using an electrical ("gate") signal of appropriate amplitude and duration at least equal to the one of the optical pulse. The "gate" is synchronized with the counter-clockwise (CCW) pulse. The pulse switched to the port T of the Sagnac interferometer is reflected by the FBG and travels again through the fiber loop mirror. At this point, no other phase shift is applied and the pulse bounces back and forth between the fiber loop and the FBG until it is extracted after a desired number of round trips applying a second electrical "gate". The pulse train is designed such that a given pulse coupled into the cavity does not overlap, inside the phase shifting element, with the subsequent pulse. As a result, the pulse entering the resonator experiences the power coefficients of the cavity in the open state ($|r_{\mathrm{FL}}|^2 = 0$ and $|t_{\mathrm{FL}}|^2 = a_{\mathrm{p}}$), while the power coefficients seen by the pulse already stored in the resonator are those of the cavity in the closed state ($|r_{\mathrm{FL}}|^2 = a_{\mathrm{p}}$ and $|t_{\mathrm{FL}}|^2 = 0$). The resonator operating in this way can be considered nonreciprocal since it exhibits unequal in-coupling and out-coupling energy rates respectively before and after the energy has been coupled in the cavity. These energy rates results to be given by the transmission coefficients of the fiber loop mirror relative to the cavity in the open and closed state respectively, divided by the cavity round trip time:

$$\rho_{FL\text{-}in} = \frac{|t_{FL}|^2_{\rightarrow}}{T_{RT}}; \quad \rho_{FL\text{-}out} = \frac{|t_{FL}|^2_{\leftarrow}}{T_{RT}}. \tag{S14}$$

In equation (S14), the arrows in the power transmission coefficients have been added to indicate the two different states: from left to right for the transmission coefficient in the open state given by $|t_{FL}|^2_{\rightarrow} = (\tau^2 + \kappa^2)^2 a_p = a_p$; and from right to left for the transmission coefficient in the closed state, which results to be $|t_{FL}|^2_{\leftarrow} = (\tau^2 - \kappa^2)^2 a_p$.

It should be noticed that, by modelling the Figure-9 cavity as a Fabry-Pérot cavity, we treat the Sagnac interferometer as a lossy mirror with attenuation factor $a_p$. However, the fiber loop mirror is itself part of the Figure-9 cavity, therefore its losses become part of the attenuation experienced by the pulse in one round trip. Therefore, if we want to find the in-coupling and out-coupling energy rates of the Figure-9 resonator, that correspond to the energy rates during the loading and decay process respectively, we have to divide the coefficients in equation (S14) by the attenuation factor $a_p$.

$$\rho_{F9\text{-}in} = \frac{1}{a_p}\frac{|t_{FL}|^2_{\rightarrow}}{T_{RT}} = \frac{\alpha_{in}}{T_{RT}} = \frac{1}{T_{RT}}; \quad \rho_{F9\text{-}out} = \frac{1}{a_p}\frac{|t_{FL}|^2_{\leftarrow}}{T_{RT}} = \frac{\alpha_{out}}{T_{RT}} = 0; \tag{S15}$$

where $\alpha_{in} = (\tau^2 + \kappa^2)^2$ and $\alpha_{out} = (\tau^2 - \kappa^2)^2$ represent the transmission coefficients of the input port of the Figure-9 resonator.

Note that, assuming an ideal coupler ($\tau^2 + \kappa^2 = 1$), $\alpha_{in}$ is always equal to 1 regardless the value of the field coupling coefficients $\tau$ and $\kappa$ of the coupler, while $\alpha_{out}$ can range from 0 to 1 according to the values of $\tau$ and $\kappa$.

By summing the coupling energy rates of the Figure-9 resonator and the energy rates due to the internal (non-radiative) losses, we obtain the loading and decay rates associated respectively to the loading and decay process used for the calculation of the TBP:

$$\rho_L = \rho_{F9-in} + \rho_0 = \frac{\alpha_{in}}{T_{RT}} + \frac{1}{\tau_0}; \quad \rho_D = \rho_{F9-out} + \rho_0 = \frac{\alpha_{out}}{T_{RT}} + \frac{1}{\tau_0}; \tag{S16}$$

**Figures:**

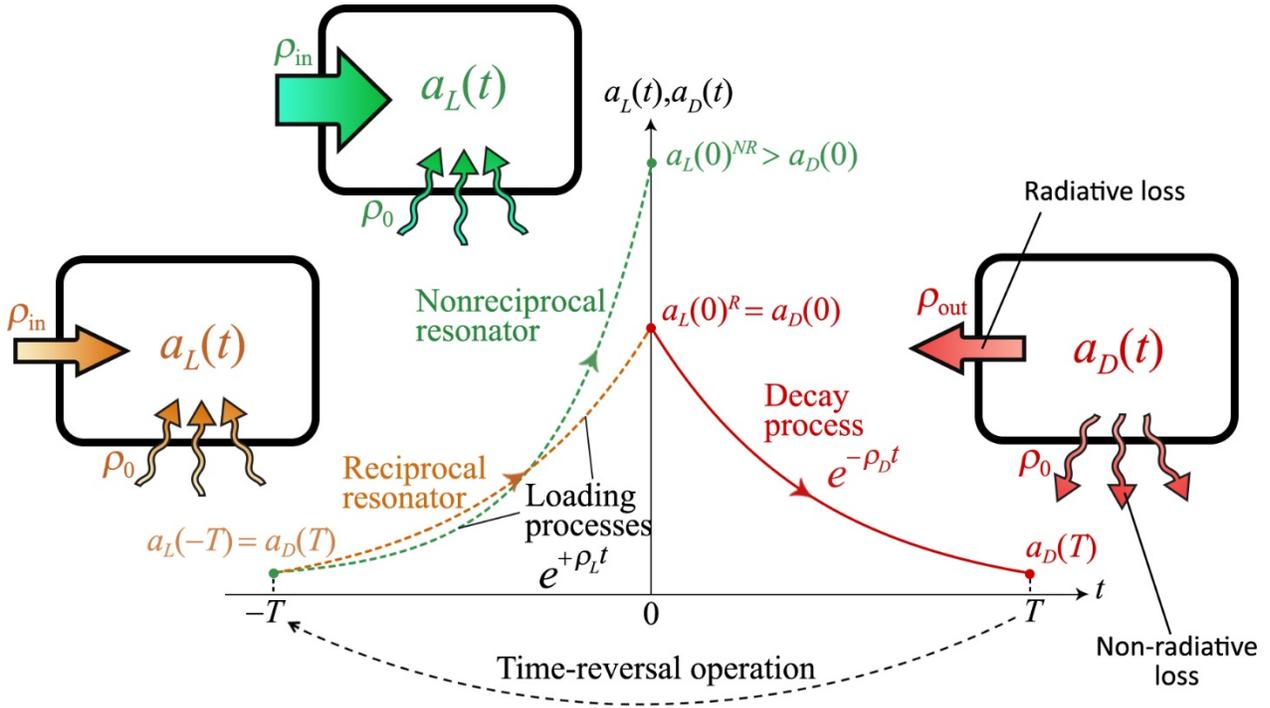

**Fig. S1.**

Graph of the time-reversal operation of the energy decay process for reciprocal and non-reciprocal resonant systems and their illustrative representations. The decay of the energy stored within the cavity (red curve) is due to the loss of power through radiative (transmission through coupling elements such as mirrors, couplers etc.) and non-radiative processes (absorption and scattering losses), which are taken into account by the out-coupling $\rho_{out}$ and intrinsic $\rho_0$ energy rate, respectively ($\rho_D = \rho_{out} + \rho_0$). The loading curves (green- and orange-dashed curves) depicts how fast the intra-cavity energy would exponentially grow if the resonator was 'fed' through the same processes reversed in time, with the radiative and non-radiative processes that becomes the in-coupling energy rate and intrinsic loading rate of energy respectively ($\rho_L = \rho_{in} + \rho_0$). If the resonant system is reciprocal, the decay process and its corresponding time-reversed process are identical. However, if the in-coupling energy rate is higher than the out-coupling energy rate, the exponential energy decay and its corresponding time-reversed process (energy loading) are different and the intra-cavity energy reaches a value at the final state greater than the initial energy state of the decay process. In this case the system is said time-reversal asymmetric, hence non-reciprocal (*22*).

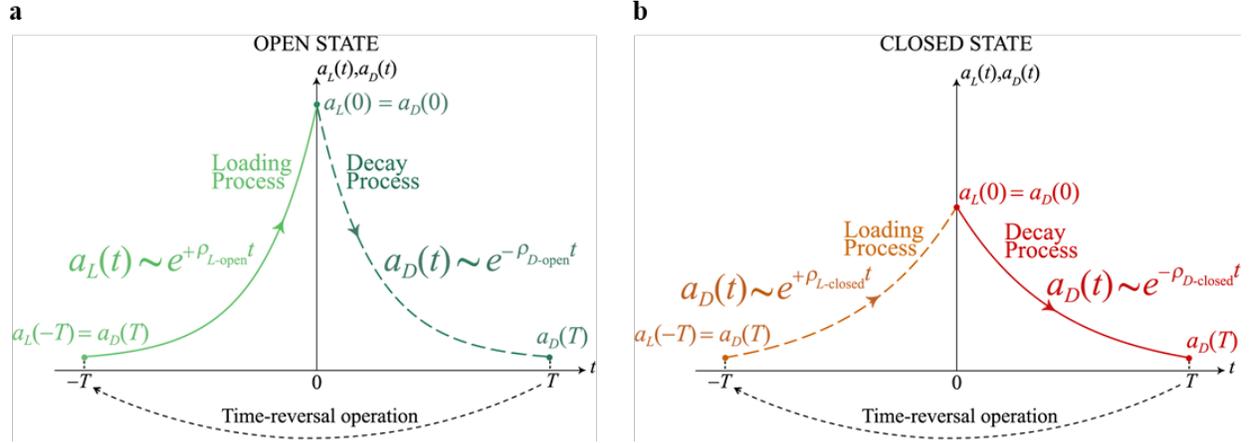

**Fig. S2.**

Loading and decay processes of a time-variant resonator. In the open state (**a**) and in the closed state (**b**) the energy in the resonant system builds up and decays at the same rate: $\rho_{L\text{-open}} = \rho_{D\text{-open}}$ and $\rho_{L\text{-closed}} = \rho_{D\text{-closed}}$ respectively. The time-reversal symmetry is never broken during each state, therefore the system results to be reciprocal at any given time. Nevertheless, the resonator is nonreciprocal owing to a breaking of its time-invariance since it switches from the open to the closed state after energy has been coupled in the resonator in a time period shorter than the round-trip time. The solid lines in **a** and **b**, represent the loading and decay process, respectively, during the open and closed state of the nonreciprocal time-variant resonator before and after the switching respectively.

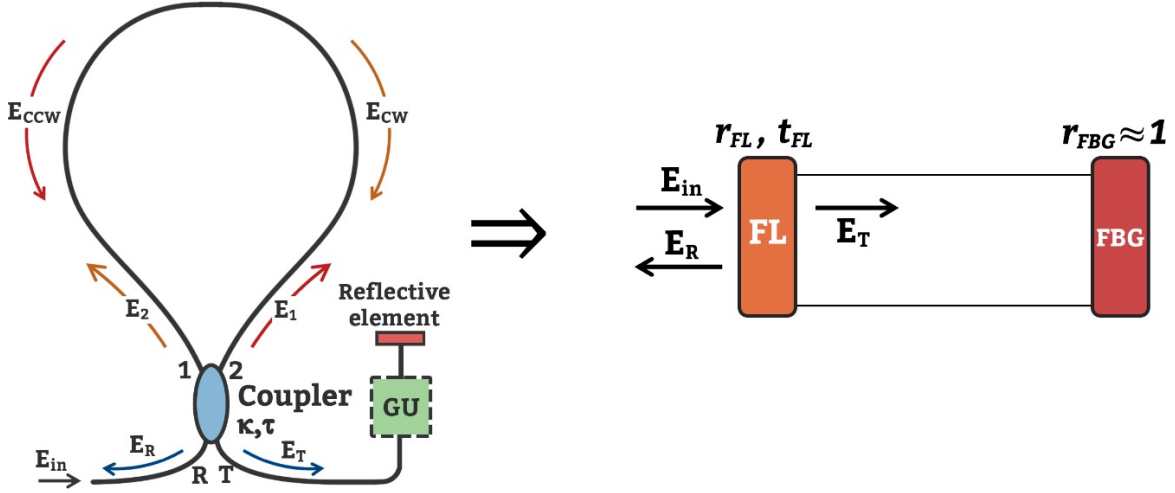

**Fig. S3.**

Basic scheme of the Figure-9 resonator (left) and the equivalent layout of a Fabry-Pérot cavity (right), where FL indicates the Sagnac interferometer (also known as a fiber loop mirror), and FBG the fiber Bragg grating.

**References**

1. Caloz, C. *et al.* Electromagnetic Nonreciprocity. *Phys. Rev. Appl.* **10**, 047001 (2018).

2. Leuchs, G. & Sondermann, M. Time-reversal symmetry in optics. *Phys. Scr.* **85**, 058101 (2012).

3. Heugel, S., Villar, A. S., Sondermann, M., Peschel, U. & Leuchs, G. On the analogy between a single atom and an optical resonator. *Laser Phys.* **20**, 100–106 (2010).

4. Mann, S. A., Sounas, D. L. & Alù, A. Nonreciprocal Cavities and the Time-Bandwidth Limit. *Optica* **6**, 104–110 (2019).

5. Tsakmakidis, K. L. *et al.* Breaking Lorentz reciprocity to overcome the time-bandwidth limit in physics and engineering. *Science* **356**, 1260–1264 (2017).

6. Sokoloff, J. P., Prucnal, P. R., Glesk, I. & Kane, M. A Terahertz Optical Asymmetric Demultiplexer (TOAD). *IEEE Photonics Technol. Lett.* **5**, 787–790 (1993).